# Dynamical evolution of the community structure of complex earthquake network

SUMIYOSHI ABE[1] and NORIKAZU SUZUKI[2]

[1] *Department of Physical Engineering, Mie University, Mie 514-8507, Japan*

[2] *College of Science and Technology, Nihon University, Chiba 274-8501, Japan*

**Abstract** — Earthquake network is known to be complex in the sense that it is scale-free, small-world, hierarchically organized and assortatively mixed. Here, the time evolution of earthquake network is analyzed around main shocks in the context of the community structure. It is found that the maximum of the modularity measure quantifying existence of communities exhibits a peculiar behavior: its maximum value stays at a large value before a main shock, suddenly drops to a small value at the main shock, and then increases to relax to a large value again relatively slowly. Thus, a main shock absorbs and merges communities to create a larger community, showing how a main shock can be characterized in the complex-network representation of seismicity.

______________________________________________________________________

Recent advances in network science have cast light on understanding physical properties of complex systems and phenomena [1,2]. They have revealed how the dynamical behavior of a complex system is governed by the underlying network structure as the system's architecture.

In ref. [3], we have introduced the concept of earthquake network, which is an evolving stochastic network constructed from a seismic time series. This network has been found to be a complex network, being scale-free [3,4], small-world [5], hierarchically organized and assortatively mixed [6]. Its heterogeneity is induced by main shocks that play roles of hubs of the network [3].

Here, we discuss the dynamical evolution of the structure of earthquake network. In particular, we focus our attention on the evolution of the community structure [7-9], that is, how the whole network is partitioned to sub-networks, in which member vertices are densely interconnected each other. Since main shocks play a role of hubs, they give an impact on the global structure of the network. In turn, it is expected that the community structure may characterize main shocks in a peculiar manner. We monitor the maximum value of the modularity measure, $Q_{\max}$, of the earthquake networks constructed from the seismic data taken from California (which is available at http://www.data.scec.org). We report the discovery of a universal behavior in the evolution of $Q_{\max}$ around main shocks. $Q_{\max}$ stays at a large value before a main shock, suddenly drops to a small values at the main shock, and then slowly increases to a large value again. Thus, main shocks are characterized in our network approach. We also make a comment on this result in connection with the clustering structure of the earthquake networks.

We start the discussion with succinctly recapitulating the method of constructing an earthquake network presented in ref. [3]. We divide a geographical region under consideration into cubic cells with the size $L \times L \times L$. A cell is regarded as a vertex of a network if earthquakes with any values of magnitude (above a certain detection threshold of seismometers) occurred therein. Then, we connect two vertices of successive events by an edge. If two successive events occur in the same cell, then a tadpole (i.e., a self-loop) is attached to that vertex. These edges and tadpoles represent the event-event correlations in seismicity. Recently, the physical basis for this representation has been clarified from the viewpoint of the concept of the internal time in seismicity termed event time [10]. The network thus constructed is a directed network with the direction from past to future and possesses multiple edges as well as tadpoles. In the small-world picture, such a full network has to be reduced to a simple undirected network. There, tadpoles are removed and each multiple is replaced by a single edge. Accordingly, the elements of the network adjacency matrix become 0 or 1 and, in particular, all diagonal elements are 0. We note that this method of constructing an earthquake network contains a single parameter, the cell size. It is important to examine how the properties of the network depend on its value. This point has recently been studied in detail in refs. [11-13]. In the present work, we shall examine two different values of the cell size: $5\,\text{km} \times 5\,\text{km} \times 5\,\text{km}$ and $10\,\text{km} \times 10\,\text{km} \times 10\,\text{km}$.

To characterize the community structure of the network, we calculate the modularity measure, $Q$, which is defined for $k$ communities as follows [7]:

$$Q = \sum_{i=1}^{k} \left( e_{ii} - a_i^2 \right), \tag{1}$$

where $e_{ij}$ stands for the fraction of all edges in the network connecting vertices in the *i*-th and *j*-th communities and $a_i = \sum_{j=1}^{k} e_{ij}$. $Q \in [0,1]$ and $Q = 0$ (1) for the absence of the community structure (the presence of the strongest community structure). We construct the dendrogram, i.e., the *hierarchical tree*, of the network and look over the hierarchies to find the maximum value, $Q_{max}$. The larger $Q_{max}$ is, the more manifest the partition of the whole network into communities is.

To detect the community structure, the concept of edge betweenness plays a key role. It is defined as follows. Take all pairs of vertices contained in a network and find the shortest path for each pair. All shortest paths between different communities pass through few edges connecting the communities. The edge betweenness is defined as the number of times passed by the shortest paths. Then, the algorithm for detecting communities proposed in ref. [7] consists of the following steps: (I) remove the edge with the highest edge betweenness, (II) recalculate edge betweenness for all remaining edges and then (III) repeat (I) and (II) until no edges remain.

In fig. 1, we present the dendrogram of the full earthquake network of seismicity for 10 days before the Hector Mine Earthquake. From the top to the bottom, the hierarchical level of the network revealed by the above-mentioned process (I)-(III) of removing the edge with the highest edge betweenness at each level of hierarchy is shown. The dots on the dead ends at the bottom are the isolated vertices. The corresponding plot of the modularity measure is given in fig. 2. We ignore the directedness since such a concept is irrelevant to the network community.

Our purpose is to monitor the value of $Q_{max}$ in the course of the evolution of earthquake network. To do so, we construct and update the earthquake network for every 10 days with no temporal overlaps. (Here, "10 days" is nothing but a daily-life scale.) As already mentioned, two different values of the cell size are examined: $5 \text{km} \times 5 \text{km} \times 5 \text{km}$ and $10 \text{km} \times 10 \text{km} \times 10 \text{km}$. We focus our attention to the celebrated three main shocks contained in the Californian data, that is, i) the Joshua Tree Earthquake (M6.1 on April 23, 1992), ii) the Landers Earthquake (M7.3 on June 28, 1992) and iii) the Hector Mine Earthquake (M7.1 on October 16, 1999).

Now, in fig. 3, we present the evolution of the maximum value of the modularity measure around those three main shocks. There, we observe a remarkable common behavior. Before the main shocks, it quasi-stationarily stays large (0.5 ~ 0.7). It suddenly drops to a small value (about 0.1) at the moments of the main shocks. And, then, it increases to return to the initial large values relatively slowly. (The quasi-stationary regime before the Landers Earthquake is actually not so clearly appreciated. This is because the time interval between the Joshua Tree Earthquake and the Landers Earthquake is very short and accordingly the relaxation from the Joshua Tree Earthquake is considered to be incomplete.) As the result, a main shock absorbs and merges communities to create a larger community, making the modularity measure smaller. Thus, we see how a main shock affects the global structure of the earthquake network.

It may be of interest to see the above result in terms of the evolution of the clustering coefficient studied in ref. [14]. The clustering coefficient (see ref. [15] for its definition) of a reduced simple earthquake network quasi-stationarily stays small before a main shock, suddenly jumps up at the moment of the main shock and then slowly decreases to become quasi-stationary again. Thus, the behavior of the modularity measure is opposite to that of the clustering coefficient. Although there is no known direct relationship between the modularity measure and clustering coefficient, this seems to be consistent if one takes into account the fact that, in a complete network (i.e., a fully connected network), the modularity measure is at the minimum, whereas the clustering coefficient is at the maximum.

In conclusion, we have studied the dynamical evolution of the community structure of the earthquake networks. We have found the universal behavior of the modularity measure around the main shocks: its maximum value quasi-stationarily stays at a large value (0.5 ~ 0.7) before the main shocks, suddenly drops to a small value (about 0.1) at the main shocks and then returns to a large value again relatively slowly. This result implies that a main shock absorbs and merges communities to create a larger community. Thus, main shocks are characterized within the network approach in a peculiar manner.

In the present work, we have evaluated the modularity measure for the full earthquake networks. Actually, we have also analyzed the simple earthquake networks reduced from the full ones and have ascertained that the same behavior is observed in the dynamical evolution of the community structure of the simple networks.

We wish to emphasize that the present result about the characteristic behavior of $Q_{max}$ around main shocks is not limited to the data in California. We have, in fact, ascertained the same behavior around the Niigata Earthquake (M6.8 on October 23, 2004) and the Kushiro-Oki Earthquake (M7.1 on November 29, 2004) in Japan. Thus, the present discovery may be universal.

*Note Added*. Following the advice of the anonymous referee, we have compared each real earthquake network with artificially generated random networks, the values of the link density of which are fixed to be identical to that of the real earthquake network. To do so, we have evaluated the *z*-score [9], which is defined by $z = (Q_{max} - \langle Q \rangle_{random}) / \sigma_Q^{random}$, where $\langle Q \rangle_{random}$ and $\sigma_Q^{random}$ are the average over 10 realizations of the modularity measure of the random networks and the associated standard deviation, respectively. The result is as follows. $z = 7.8318$ for the earthquake network for 10 days before the Hector Mine Earthquake and $z = -8.5206$ for 10 days after the Hector Mine Earthquake. This shows how the community structure of the real earthquake networks is different from that of the random networks.

\* \* \*

SA was supported in part by a Grant-in-Aid for Scientific Research from the Japan Society for the Promotion of Science.

# Figure Caption

Fig. 1  The dendrogram of the earthquake network constructed from the seismic time series between October 6, 1999 and October 16, 1999. The cell size taken is $5\text{km} \times 5\text{km} \times 5\text{km}$.

Fig. 2  The plot of the modularity measure with respect to the level of hierarchy associated with the dendrogram in fig. 1. $Q_{max} = 0.682....$

Fig. 3  The evolution of the maximum value of the modularity measure, $Q_{max}$, around i) the Joshua Tree Earthquake, ii) the Landers Earthquake and iii) the Hector Mine Earthquake. The moments of these main shocks are located at the origins. In each case, the values of the cell size are: a) $5\text{km} \times 5\text{km} \times 5\text{km}$ and b) $10\text{km} \times 10\text{km} \times 10\text{km}$.

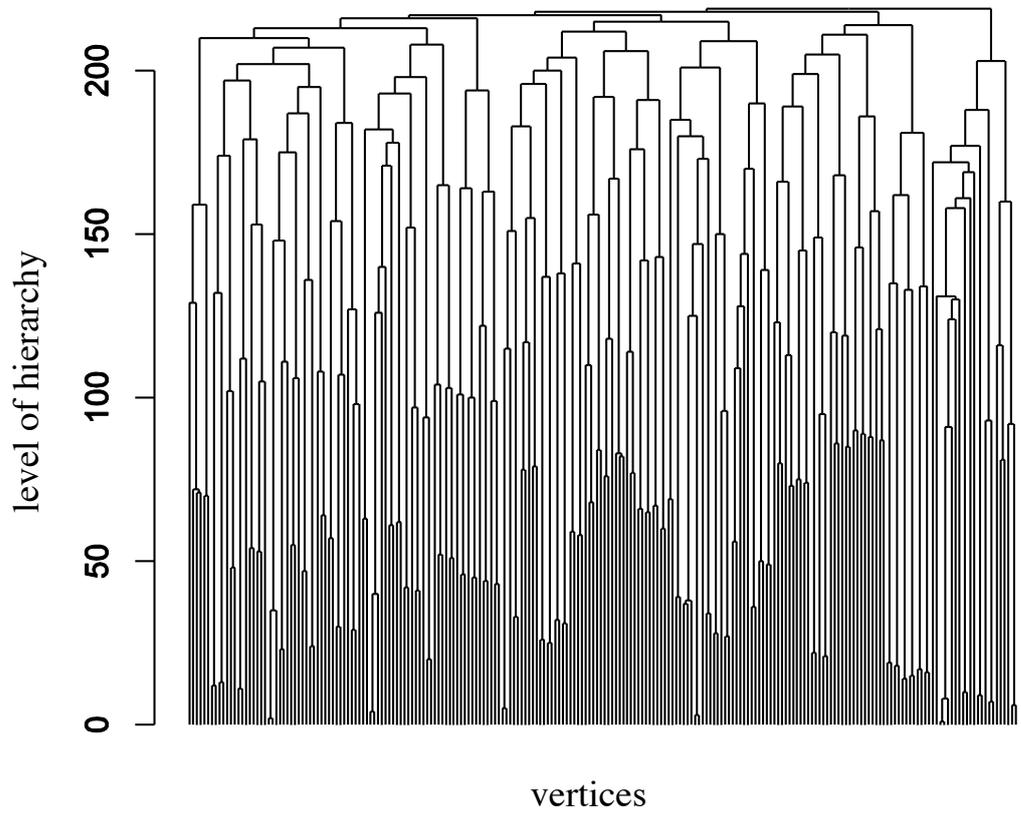

Fig. 1

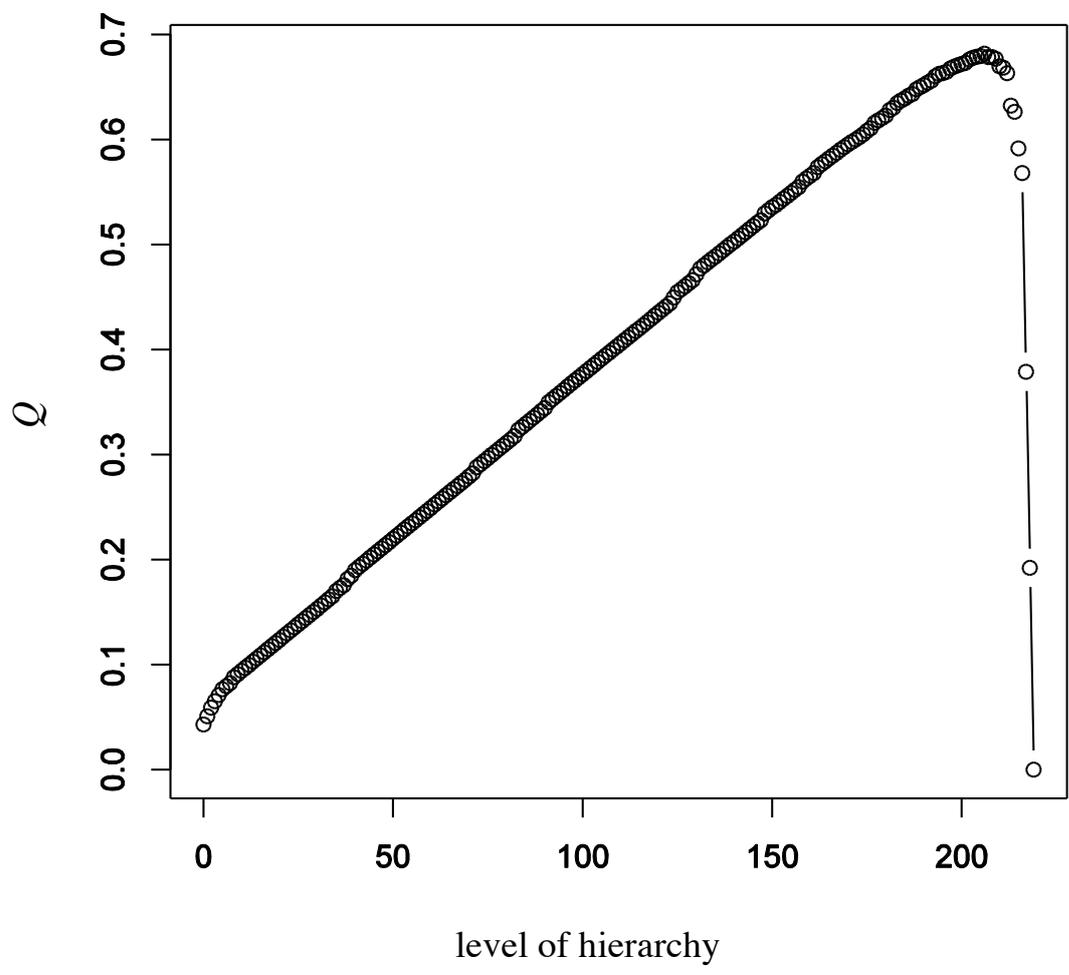

Fig. 2

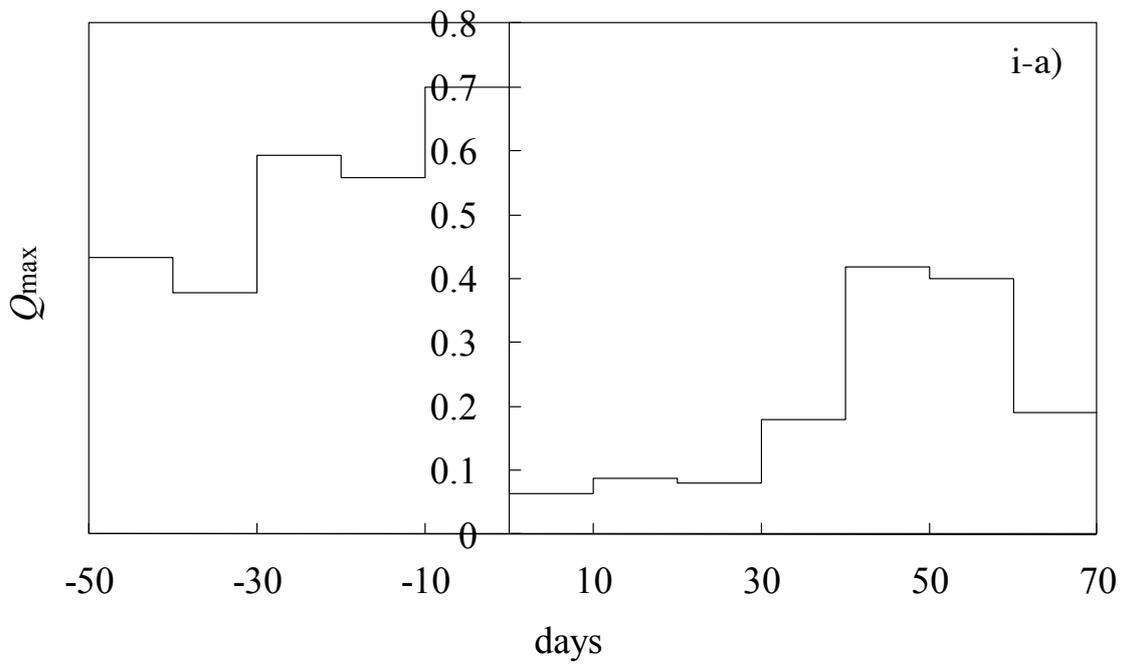

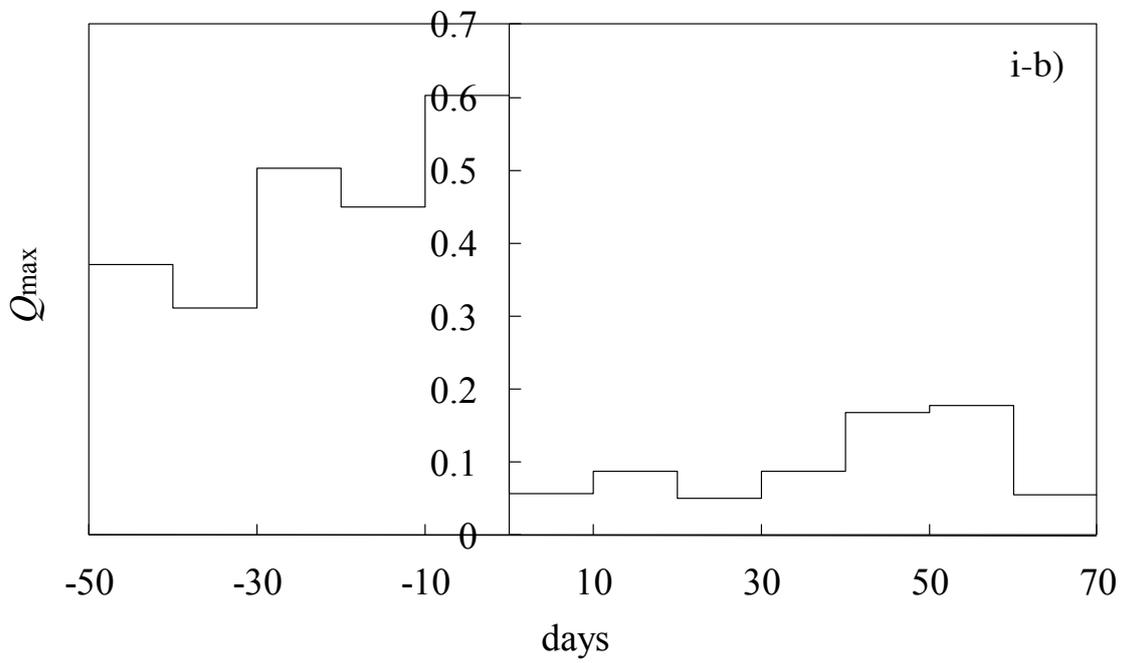

Fig. 3 i)

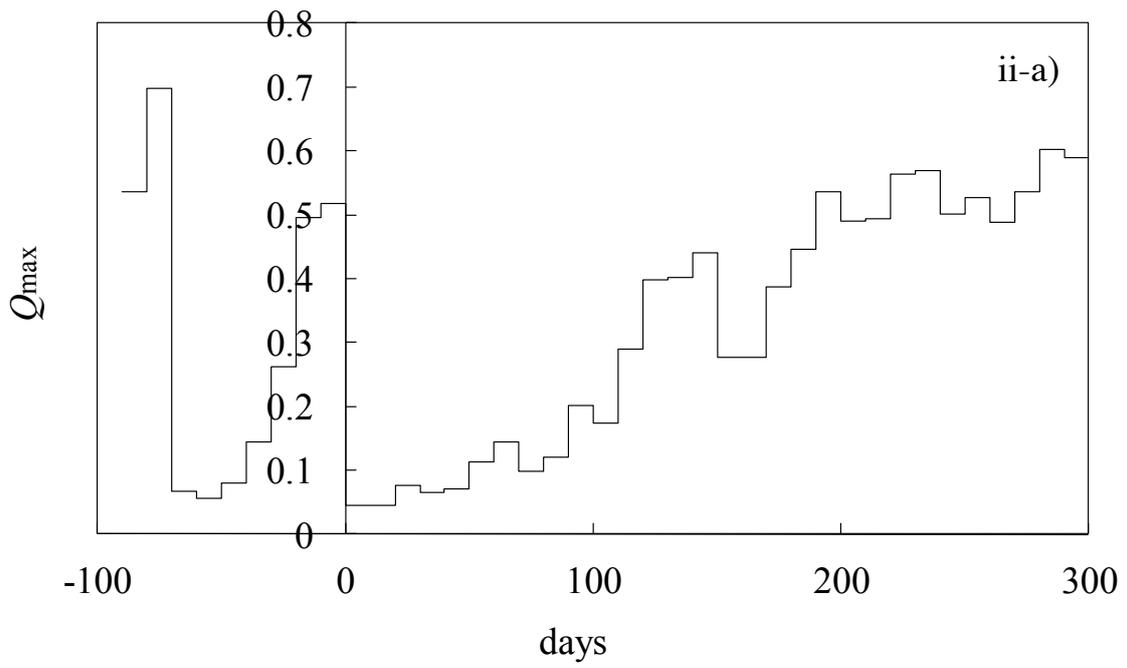

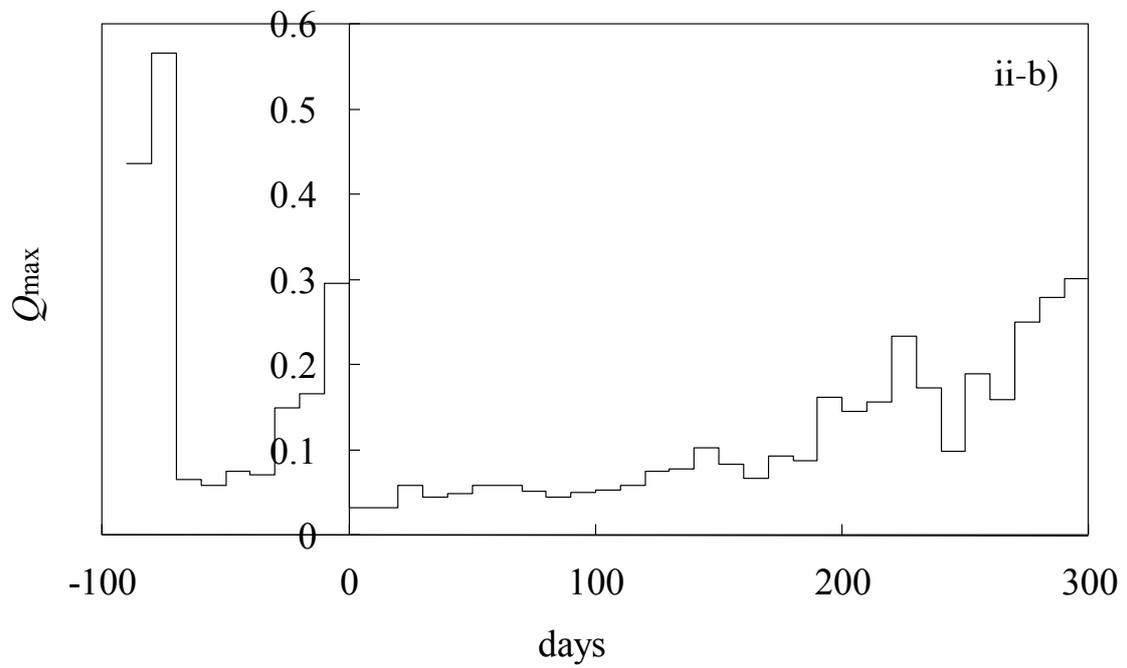

Fig. 3 ii)

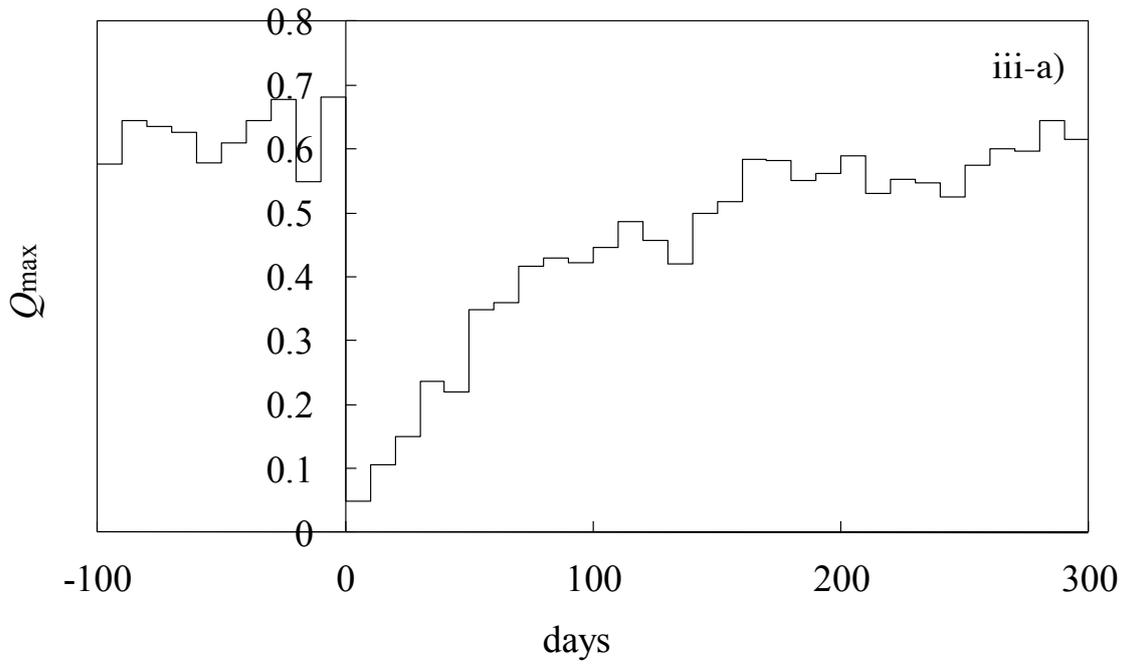

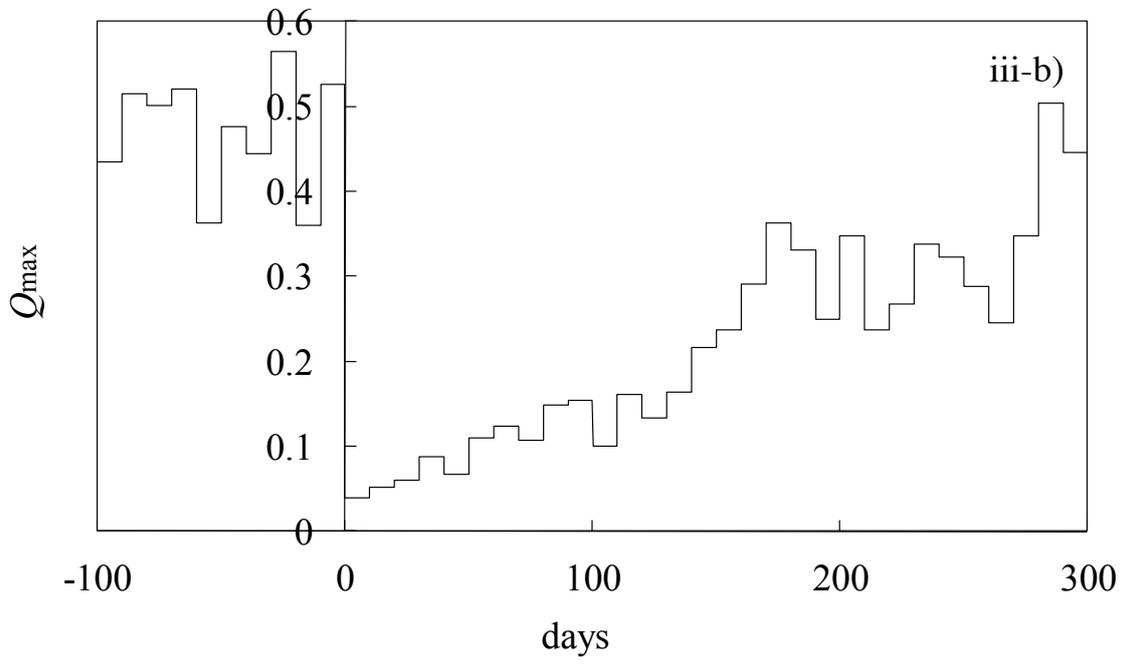

Fig. 3 iii)